\documentclass[manuscript]{aastex62}

\usepackage{natbib}
\usepackage{graphicx, epsfig, hyperref}
\def\up#1{\leavevmode \raise.16ex\hbox{#1}}

\newcommand\kms{${\rm km\,s}^{-1}$}
\newcommand\oiline{\up[\ion{O}{1}\up]\,$63.19\,\mu$m~}
\newcommand\siline{\up[\ion{S}{1}\up]\,$25.25\,\mu$m~}
\newcommand\feiiline{\up[\ion{Fe}{2}\up]\,$25.99\,\mu$m~}

\newcommand\sofias{{\it SOFIA~}}
\newcommand\sofia{{\it SOFIA}}

\received{2020 March 18}
\revised{2020 March 25}
\accepted{}

\submitjournal{ApJL}

\shorttitle{{\it SOFIA}-EXES Observations of Betelgeuse During the Great  Dimming}
\shortauthors{Harper et al.}

\begin{document}

\title{{\it SOFIA}-EXES Observations of Betelgeuse during the Great Dimming of 2019/2020}

\correspondingauthor{Graham Harper}
\email{graham.harper@colorado.edu}

\author[0000-0002-7042-4541]{Graham M. Harper}
\affiliation{Center for Astrophysics and Space Astronomy\\
University of Colorado Boulder \\
389 UCB\\
Boulder, CO 80309\\
USA}

\author{Curtis N. DeWitt}
\affiliation{USRA/SOFIA\\
NASA Ames Research Center\\
Moffett Field, CA 94-35\\
USA}

\author{Matthew J. Richter}
\affiliation{Department of Physics\\
University of California Davis\\ 
Davis, CA 95616\\
USA}

\author{Edward F. Guinan}
\affiliation{Astrophysics and Planetary Science Department\\
Villanova University\\
Villanova, PA 19085\\
USA}

\author{Richard Wasatonic}
\affil{Astrophysics and Planetary Science Department\\
Villanova University\\
Villanova, PA 19085\\
USA}

\author{Nils Ryde}
\affiliation{Department of Astronomy and Theoretical Physics\\
Lund University\\
Lund\\
Sweden}

\author{Edward J. Montiel}
\affiliation{USRA/SOFIA\\
NASA Ames Research Center\\
Moffett Field, CA 94-35\\
USA}

\author{Amanda J. Townsend}
\affiliation{Department of Physics\\
University of California Davis\\ 
Davis, CA 95616\\
USA}

\begin{abstract}

In 2019 October Betelgeuse began a decline in V-band brightness that went beyond the minimum 
expected from its quasi-periodic $\sim 420$\,day cycle, becoming the faintest in recorded photometric history.
Observations obtained in 2019 December with VLT/SPHERE (Montarg\`es 2020) have shown that the southern half of the star
has become markedly fainter than in 2019 January indicating that a major change has occurred in, or near, the
photosphere.

We present {\it SOFIA}-EXES high spectral-resolution observations of \feiiline and \siline emission lines
from Betelgeuse obtained during the unprecedented 2020 February V-band brightness minimum to investigate
potential changes in the circumstellar flow. These spectra are
compared to observations obtained in 2015 and 2017 when the V magnitude was typical of brighter phases.
We find only very small changes in the gas velocities reflected by either of the line
profiles, no significant changes in the flux to continuum ratios, and hence no significant changes in the \up[\ion{Fe}{2}\up]/\up[\ion{S}{1}\up] flux ratios. There is evidence that absorption features have appeared in the 2020 continuum.

The Alfv\'en wave-crossing time from the upper-photosphere is sufficiently long that one would not expect
a change in the large scale magnetic field to reach the circumstellar
\up[\ion{Fe}{2}\up] and \up[\ion{S}{1}\up] line forming regions, $3 < R (R_\ast) <20$. However, 
the light-crossing time is of order a few hours and a reduction in luminosity
may reduce the dust-gas heating rate and \oiline emission which has contributions from $R> 20 R_\ast$, where significant circumstellar oxygen-rich dust is observed.

\end{abstract}


\section{Introduction} \label{sec:intro}

Betelgeuse ($\alpha$~Orionis, HD 39801, HR~2061)  is a massive nearby red supergiant (RSG) which is destined to undergo core-collapse to become either a supernova or implode directly into a black hole. RSGs and supernovae play an important role in the chemical evolution and shaping of galaxies but, despite this, the RSG phase of evolution is not well understood. This is, in part, because the stellar evolution depends on both mass loss and rotation, which are not well characterized
\citep[e.g.,][]{Meynet_2015A&A...575A..60M}. Furthermore there is no {\it standard model} that can predict mass-loss rates from RSGs and which can explain the panoply of observed stellar outflow diagnostics. This problem is particularly acute for K and early M supergiants, including well-known stars such as Betelgeuse and Antares, because their outflows are under-abundant in molecules and dust compared to their later-spectral type counterparts, where radiation pressure on molecules and/or dust may drive outflows.

Betelgeuse's large angular size and brightness have resulted in it being extensively studied at multiple wavelengths over the last century,  with some recent studies presented in 
\citet{BW201_2013EAS....60.....K}. These studies seek to unravel the puzzling origins of different species of dust seen at different radii, the extended photosphere and chromosphere, quasi-steady outflows, photospheric hot spots, and atmospheric asymmetries.

Early photometric \citep{Stebbins_1931PWasO..15..177S}
and radial velocity studies \citep{Spencer_1928MNRAS..88..660S, Sanford_1933ApJ....77..110S} both revealed a period of 
$\simeq 5.8$\,yr ($\simeq 2100$\,day) 
with peak-to-peak variations of 0.44 magnitude, and $\simeq 4-6$\,\kms, respectively. A discussion of these data and observations over the next few  decades is given by \citet{Goldberg_1984PASP...96..366G}. In addition to shorter ($\sim$\,week)  time-scale fluctuations, the early photometric data also showed indications of a $\sim 420$\,day period later discovered in the satellite ultraviolet and B magnitude data by \cite{Dupree_1987ApJ...317L..85D}, and confirmed in the radial velocity study of \citet{Myron_1989AJ.....98.2233S}. Betelgeuse has an 
MK spectral-type of M1-M2~Ia-ab \citep{KM1989ApJS...71..245K}, but it also exhibits a variable spectral-type;
\citet{WW1978ApJ...222..209W}  using narrow band photometric TiO and CN indices derived 
a mean spectral-type of M2.2 with a range of M1.5-M2.7 between 1969 and 1976. 
Despite the lack of a clear constant phase relation between the radial velocity and light curve data
the variations probably result from the interaction of convection in the outer envelope and pulsation \citep{Kiss_2006MNRAS.372.1721K}.  How these photospheric variations are connected to the heating of the extended atmosphere and circumstellar envelope (CSE), and the ejection of mass is not known. It is of particular interest given the importance of RSGs as Type II supernova progenitors, e.g., Betelgeuse, with an initial mass of $\sim20 {\rm M}_\odot$ \citep{Dolan_2016ApJ...819....7D,Harper_2008AJ....135.1430H}, continuing its currently mass-loss rate might eventually undergo core-collapse to become either a Type II-P supernova \citep{Smith_2009AJ....137.3558S} or implode directly into a black hole
\citep{Smartt_2009ARA&A..47...63S,Adams_2020MNRAS.492.2578S}.
While parts of the chromosphere show evidence of non-radial motions \citep{Lobel_2001ApJ...558..815L}, the circumstellar  outflow velocities appear dynamically decoupled from the photospheric radial velocity variations \citep{Weymann_1962ApJ...136..844W}.  

The CSE consists of a dominant gas phase of neutral and singly ionized atomic species, together with under-associated CO and oxygen-rich dust. For Betelgeuse, beyond the 
extended chromosphere at $\simeq 1.75R_\ast$ \citep{Eamon_2015A&A...580A.101O},
there are two well determined outflows, an inner flow of 
$\simeq 9$\,\kms~ that extends out to $\sim 4$\arcsec, and an outer flow of 16\,\kms~that extends beyond 32\arcsec~\citep[][and references therein]{Eamon_2012AJ....144...36O}.
$7.76-19.50\>\mu$m imaging with VLT/VISIR reveals dust emission from an irregular ring-like structure between $0.5-1.0$\arcsec\,  ($23-45\>R_\ast$, assuming a photospheric angular diameter of $\phi=44$\,mas) from the star, and more diffuse irregular emission out to 3\arcsec~\citep[][and references therein]{Kervella_2011A&A...531A.117K}.  Three bright plumes also appear to extend in towards the star and two of these correspond to near-photospheric features seen in VLT/NACO JHK images \citep{Kervella_2009A&A...504..115K}.
Closer to the star, dust has also been detected in a shell near $3R_\ast$ in VLT/SPHERE/ZIMPOL polarization images \citep{Kervella_2016A&A...585A..28K}. How the 
gas and dust are mixed is not known but where they co-exist dust grains will be
driven by the stellar radiation field and heat the gas through collisions.

\begin{figure}[th]
\begin{center}
\epsscale{1.00}
\epsfig{file=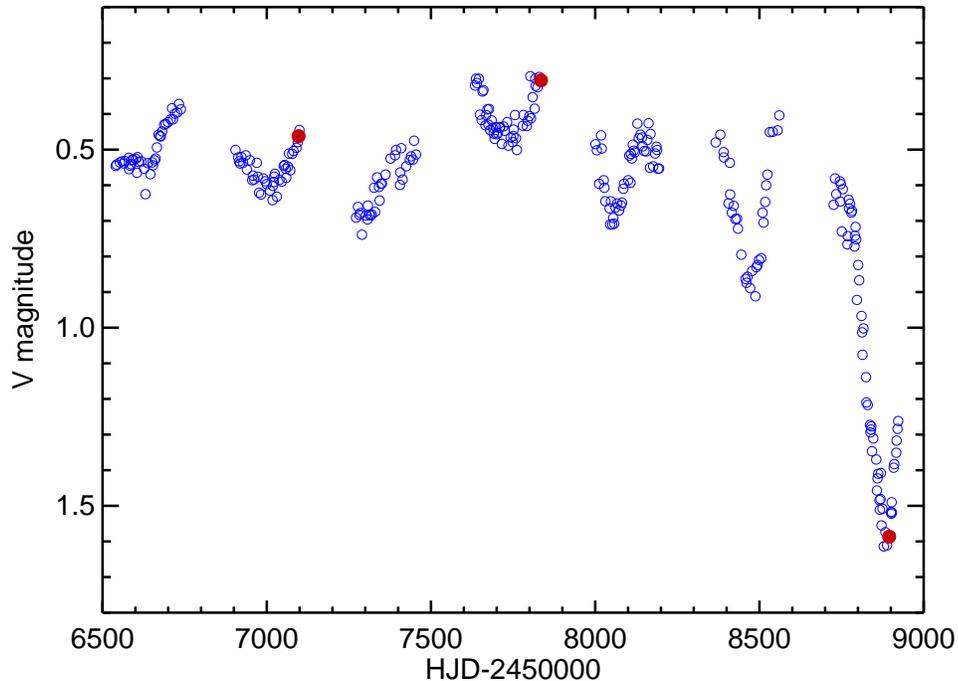, angle=90., width=13.5cm}
\caption{V magnitude photometry of Betelgeuse (blue circles) obtained at the Wasatonic Observatory, Villanova University. These data clearly show the quasi-periodic $\sim 420$\, day variations, and periodogram analyses reveal the longer 5.8\,yr period. The amplitudes and periods all appear to change.  Also shown are the magnitudes at the 2015, 2017, and 2020 epochs of the EXES observations (red-filled). It can be seen that the Cycle 2 (2015) and Cycle 5 (2017) spectra were obtained in brighter phases, while the Cycle 7 DDT observations were made at the minimum of the 2019/2020 dimming.}
\label{fig:wasatonic}
\end{center}
\end{figure}

The dynamical time-scale for motions, shocks or magnetic waves to reach the circumstellar outflow are of the order of years to decades, but  \citet{Haas1995ASPC...73..397H} observed that \oiline emission fluxes, formed between $3-100R_\ast$, might be responsive to changes in the photospheric V magnitude. They noted that two observations made with the
Cryogenic Grating Spectrometer (CGS) on the {\it Kuiper Airborne Observatory (KAO)} taken 22 months apart, on 1992 January 16 ($F=2.4\pm 0.2\>{\rm W\>cm}^{-2}$, V$\sim 0.35$) 
\citep{Haas_1993ApJ...410L.111H}
and 1993 November
($F=1.1\pm 0.2\>{\rm W\>cm}^{-2}$, V$\sim 0.59$), showed a factor of two decrease in flux when
the V-mag increased by 0.24, and an observation with 
the Far-Infrared Imaging Fabry-Perot Interferometer (FIFI) on {\it KAO} in 1993 March gave a non-detection with an upper-limit of $F(3\sigma) < 0.6\>{\rm W\>cm}^{-2}$ (V$\sim 0.88$) when the V-mag was 0.29-mag higher again.  More recently \citet{Castro-Carrizo2001A&A...367..674C} reported a 1997 September 12 {\it ISO} LWS02 flux of
$F=1.93\pm 0.06\>{\rm W\>cm}^{-2}$ when V$\sim 0.57$. The V magnitudes near the times of these observations are from \citet{Kri_1992IBVS.3728....1K, Kri_1994IBVS.4028....1K} and Fig.~\ref{fig:wasatonic}.  If there is a causal connection between V magnitudes and \oiline flux then it must be related to changes in the radiation interacting with dust in the
gaseous outflow where the flux is emitted, and which can occur on time-scales of a few hours.  For
dust-gas heating the \oiline fluxes related to changes in stellar luminosity before the observations were obtained. We also note that \citet{Skinner_1997MNRAS.288..295S}  reported a decrease in the surface brightness of UKIRT 9.7\,$\mu$m and 12.5$\mu$m CSE silicate images, extending out to $3-4 \arcsec$, over a 1-yr interval during the time when there was a sudden decrease in V-band brightness \citep{Guinan_1993IAUC.5708....1G}.

Since 2019 October  Betelgeuse has undergone a deep decline of over 1.1-mag~ 
\citep{Guinan_2020ATel13512....1G}, exceeding the previous quasi-periodic minimum of $V\simeq 0.9$\,mag, and  becoming the faintest it has been in modern record at V=1.61\,mag during
2020 February 07-13. Fig.~\ref{fig:wasatonic} shows V photometry obtained over a span of the last 23-years obtained at the Wasatonic Observatory, Villanova University where it can be seen that
while the most recent minimum has occurred close in time to that expected based on previous cycles,
its depth is unprecedented. Observations obtained in 2019 December with the VLT/SPHERE in the CntHa filter
($\lambda644.9, \Delta\lambda=4.1$\,nm)
(Montarg\`es 2020\footnote{\href{https://www.eso.org/public/images/eso2003c/}{ESO Press Release}}  , Priv. Comm.) have shown that the southern half of the photosphere has become markedly fainter than in 2019 January. 
Potential causes for the deep decline beyond the typical minimum are a continued cooling of the photosphere leading to deep TiO absorption bands in V-band and/or  the formation of dust obscuring the photosphere.
\cite{Levesque_2020ApJ...891L..37L} 
find that the mean $T_{eff}$ has declined slightly but this is not sufficient to explain
the optical dimming, and they propose that the presence of new large-grain dust in the line-of-sight is a possible explanation for the recent photometric changes.

Betelgeuse is currently being observed as part of an ongoing 2019/2020 CfA MOB program and with the continued decline of the V brightness reported by \citet{Guinan_2020ATel13439....1G}  astronomers have been actively studying the response of Betelgeuse with multi-wavelength observatories.  Based on community inputs of the potential scientific importance of the Betelgeuse event, the Director of Science Mission Operations for NASA-DLR's {\it Stratospheric Observatory for Infrared Astronomy (SOFIA)}
 initiated a set of DDT observations of the star in winter/spring 2020.  These observations have, or will, cover all the instruments scheduled to fly on the airplane in the period in question (EXES, FIFI-LS, upGREAT and FORCAST).  

Here we present \sofia-EXES \feiiline and \siline emission line spectra that form part of this campaign as a concerted effort to gain empirical constraints on the unprecedented dimming of
Betelgeuse.  These diagnostics are described in Table~\ref{tab:atomic}.
The EXES observations presented here provide an examination of the
line forming region, $2 < R_\ast ({\rm R}_\odot) < 20$ between the photosphere and overlapping the \oiline forming region,  $2 < R_\ast ({\rm R}_\odot) < 130$  
 \citep{Harper_2009ApJ...701.1464H}.

The EXES observations and line profile measurements are presented and discussed in 
\S\ref{section:observations} and \S\ref{section:results}, respectively. The results are analyzed and the implications for potential changes of  \oiline are given in \S\ref{section:discussion}, and the conclusions are given in \S\ref{section:conclusions}.

\section{\sofias EXES Observations\label{section:observations}}

EXES - the Echelon Cross Echelle Spectrograph - provides spectral resolutions up to R=50,000-100,000 and features a $1024\times 1024$ Si:As detector array that covers 4.5 to 28.3 microns \citep{EXES_2018JAI.....740013R}.  It is a PI class instrument that is flown on board \sofias \citep{SOFIA_2012ApJ...749L..17Y}.
In observing  Cycle 7, during the period when Betelgeuse reached its V-band brightness minimum, EXES was mounted on \sofias and observations of the circumstellar \feiiline and \siline emission lines were obtained with the same spectrograph settings as previous observations made in Cycle 5.  The \feiiline had first been observed with a different setting in Cycle 2
\citep{Harper_2017ApJ...836...22H}. The times of the observations are given in Table~\ref{tab:observations}, along with 
the V magnitudes interpolated from the light curve shown in Fig.~\ref{fig:wasatonic}. The Cycle 2 and 5 observations were obtained when Betelgeuse was in its normal bright quasi-period state, while the Cycle 7 observation was taken at its minimum.

Two slits were used for the observations which were nodded along the 28.5\arcsec ~long apertures. The default slit width is 3.23\arcsec, with a resolution of $R\simeq 65,000$, and a narrow 0.81\arcsec ~slit  which provides $R\simeq 85,000$. For Cycles 5 and 7 the \feiiline was observed through the narrow slit
while the \siline and Cycle 2 \feiiline were observed through the default slit.  At these wavelengths and high spectral-resolution there are gaps between the orders. In Cycle 5 the telluric calibrator Metis was also observed to provide the shape of the individual order sensitivities.  A measure of the image quality for {\it SOFIA} which includes telescope diffraction and pointing jitter is $\sim 4$\arcsec~FWHM, i.e.
it is wider than the slit widths but smaller than the nod distance.
The wavelength scales were derived from emission line obtained in adjacent orders, and
for Cycle 5 and 7 the uncertainty is expected to be $\simeq 0.5$\kms. Hereafter, we refer to
Cycles 2, 5, and 7 as CY02, CY05, and CY07.

\begin{table}[th]
\begin{center}
\caption{EXES Forbidden Line Diagnostic Transitions}
\begin{tabular}{clcccc}\hline
Species & Transition & Wavenumber & Wavelength  & $E_{up}$ & Einstein A-value \\
        & Upper$\to$Low & $({\rm cm}^{-1})$  & $(\mu {\rm m})$  & (K)  & $({\rm s}^{-1})$  \\
\up[Fe~II\up] & $^6D_{7/2}- {}^6D_{9/2}$  & 384.7872   & 25.98839  & 550  & $2.13\times 10^{-3}$ \\
\up[S~I\up]   & $^3P_1 - {}^3P_2 $      & 396.0587    & 25.24878    & 570  & $1.40\times 10^{-3}$  \\\hline
\end{tabular}
\tablecomments{Einstein A-values for \feiiline and \siline are from \cite{Bautista_2015ApJ...808..174B} and \citet{FF_2006ADNDT..92..607F}, respectively. The wavelengths are taken from the energy levels of \citet{Nave_2013ApJS..204....1N} and \citet{Blondel_2006JPhB...39.1409B}, respectively.}
\label{tab:atomic}
\end{center}
\end{table}

\section{Results\label{section:results}}

\begin{table}
\begin{center}
\caption{\sofia-EXES Observations of Betelgeuse. UT time of mid-observation.}
\label{tab:observations}
\begin{tabular}{lccc}
\tableline\tableline
Program ID               &    \up[Fe~II\up] &   \up[S~I\up]  & V-magnitude\\
                                & \multicolumn{2}{c}{UT yr-mon-day hh:min}\\ \tableline
Cycle 2  02$\_$004   &  2015-03-03  15:19    &  \nodata   &  0.46 \\ 
Cycle 5  05$\_$007   &  2017-03-22  06:16    &  2017-03-22  06:50  &  0.31 \\
Cycle 7  75$\_$005   & 2020-02-15   04:51    &   2020-02-15  05:09  & 1.59\\
\tableline
\end{tabular}
\end{center}
\end{table}

\begin{figure}[th]
\begin{center}
\epsfig{file=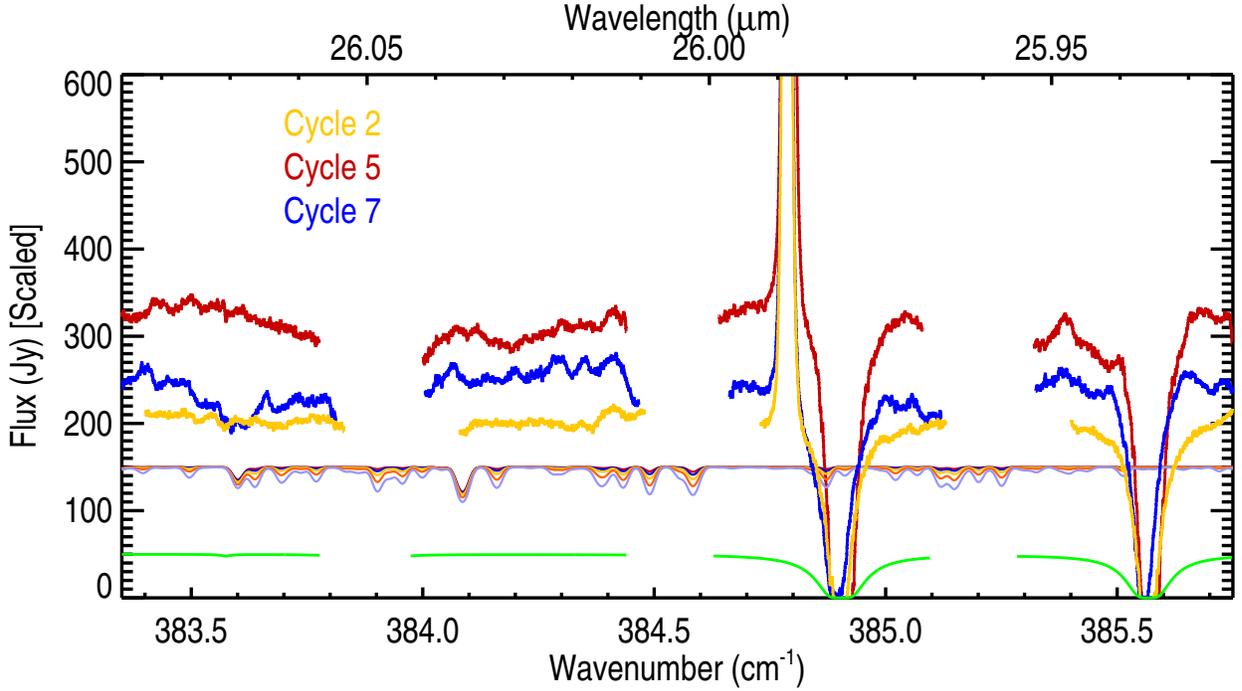, angle=90., width=18.0cm}
\caption{The spectral orders near the \feiiline line. The bottom green line is the sky spectrum which
shows the two strong telluric features. The spherical MARCS model synthetic photospheric spectra are shown below the observed spectra, normalized to 150 Jy, show the change in depth of photospheric absorption features as the effective temperature declines. The curves are, from the top downwards, for $T_{eff}=3600, 3500, 3400, 3300, 3200$\,K.}
\label{fig:orders_feii}
\end{center}
\end{figure}

The spectral orders for the \feiiline setting are shown in Fig.~\ref{fig:orders_feii}. In Cycle 7  the order shapes  appear relatively flat, while for Cycle 5  we have used the telluric calibrator Metis to apply a polynomial correction to the continuum shape. There are noticeable differences in 
the shape of the in-order continua, especially the small-scale features present in Cycle 7 which
have characteristic scales of $\sim 15$\,\kms.  To compare the observed continua with theoretical models we have computed synthetic spectra from a grid of spherical MARCS photospheric models \citep{MARCS_2008A&A...486..951G}.  The models have $T_{eff}=3600, 3500, 3400, 3300, 3200$\,K, $Log(g_\ast)=0.0$, solar abundances, except for a lower $C^{12}/C^{13}$ ratio typical of red giants. The synthetic spectra have been convolved with a $v_{macro}=15$\kms. For this spectral region, there is perhaps some correspondence of the absorption features in CY07, however the MARCS models are not a good match at any of the epochs.  The star's continuum at 25.99$\mu$m has contributions from the photosphere and optically-thin olivine silicate dust emission which dominates at $\lambda > 17\>\mu$m  \citep{Verhoelst_2006A&A...447..311V}. It is also likely that the continuum has a component from the molecular reservoir  located between the photosphere and chromosphere
\citep{Tsuji_2006ApJ...645.1448T,Perrin_2007A&A...474..599P,Montarges_2014A&A...572A..17M}. Under these circumstances it is reasonable to expect that there will be some mismatch between the observed and MARCS synthetic spectra.

\begin{figure}[th]
\begin{center}
\epsscale{0.85}
\epsfig{file=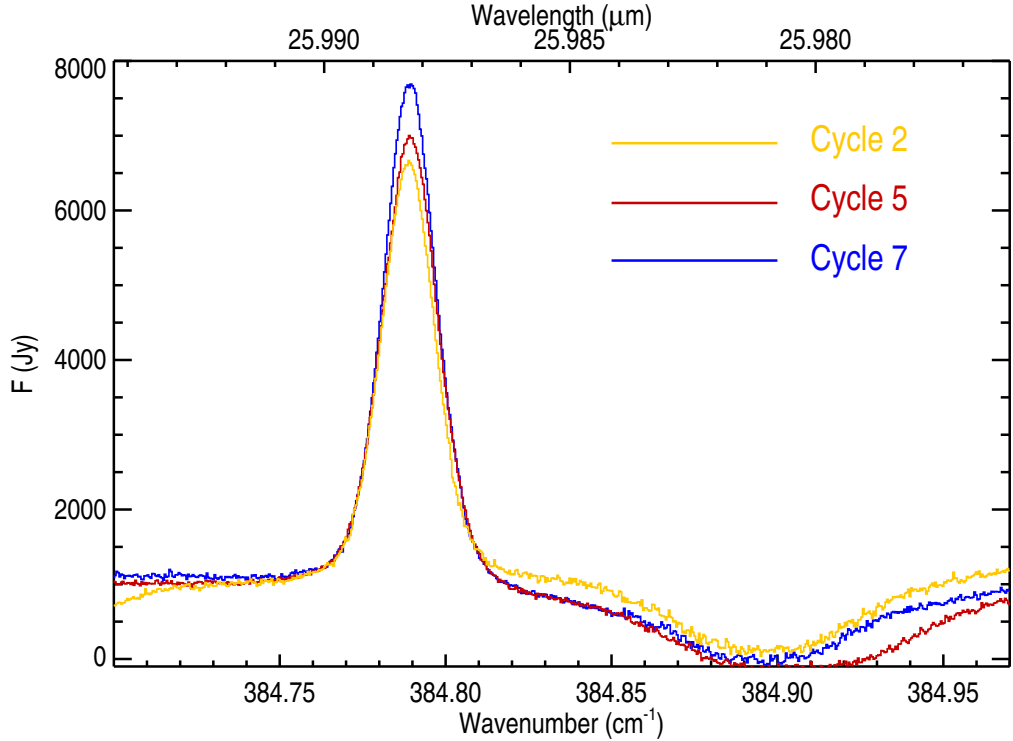, angle=90., width=13.5cm}
\caption{The Cycle 2, 5 and 7 \feiiline profiles, with CY05 and CY07 obtained with the same spectrograph setup. The profiles are centered on the stellar rest-frame, except an additional shift of 2.6\,\kms, has been applied to CY07. The continua have been normalized at 384.75\,${\rm cm}^{-1}$, close to the emission line. Note the very close agreement between the line and continuum in Cy05 and Cy07.  Note that the emission lines sit upon the far wing of the very strong telluric feature at  $384.90\>{\rm cm}^{-1}$.}
\label{fig:feii}
\end{center}
\end{figure}

The CY02, CY05 and CY07 \feiiline emission line profiles are shown in Fig.~\ref{fig:feii} where we have normalized the continuum at 384.75\,${\rm cm}^{-1}$ to 1100\,Jy, based on the
{\it IRAS} 25\,$\mu$m PSC flux \citep{IRAS_1988iras....7.....H}, color-corrected with the {\it ISO}-SWS spectrum \citep{Justtanont_1999AA...345..605J, Sloan2003ApJS..147..379S}. It is remarkable how similar the line to continuum flux ratios are for these epochs. 

To compare the \feiiline line properties, a Voigt profile and low order polynomial were used  to model the strong telluric feature near $384.90\>{\rm cm}^{-1}$.  We fit simple Gaussian profiles to the emission lines which provide a very good fit although there is a small excess in the wings.
Shown in this way the differential properties are more clearly seen. The integrated \feiiline  flux to the local continuum flux ratios for CY02:CY05:CY07 are 
0.85:1.00:1.04.  The CY05 and CY07 ratios are very similar to each other and were observed with the same settings, while the CY02 observation used a wider
slit which would lead to slightly more CSE silicate dust continuum emission passing through the aperture and potentially reducing the ratio.  For the centroid velocities we adopt a center-of-mass 
(C-o-M) radial velocity of Betelgeuse of 
$V_{rad}=20.9 \pm 0.3$ \citep{Harper_2017ApJ...836...22H} which is in good agreement with the more recent value of 
$V_{rad}=20.7\pm 0.2$\,\kms\, derived from modeling of spatially resolved stellar rotation in  SiO and CO spectra measured with {\it ALMA} by \citet{Kervella_2018AA...609A..67K}.  Observations of the photospheric radial velocity have been made during the V-band dimming and
the velocity of the photosphere has increased by 10\,\kms~ moving from blue-shifted to red-shifted
(T. Granzer \& K. Strassmeier, Priv Comm 2020). At the time of the EXES CY07 observation
$V_{phot}\simeq+2.6$\,\kms~ in the C-o-M frame.

The line profile measurements of the \feiiline are given in Table~\ref{tab:results}. The velocity centroids and line widths measured in this similar manner show no major changes\footnote{The line width of CY02 is slightly larger than reported previously because of the different telluric correction model.}. The uncertainties given here are based on the line fits and changes resulting from different continuum shape models. The difference in velocity centroids between CY05 and CY07 may be real because the observations used the same spectral settings and wavelength calibration. This could indicate a slight reduction in outflow velocity in the line forming region, however, differences in the strong telluric feature and stellar continuum shape must be considered.

\begin{table}
\begin{center}
\caption{EXES Line Profile Measurements. All velocity measurements in \kms.}
\label{tab:results}
\begin{tabular}{ccccc}
\tableline
Observation Date  &  \multicolumn{2}{c}{\up[Fe~II\up]} &   \multicolumn{2}{c}{\up[S~I\up]} \\
                            &   $V_{cent}$  & $V_{FWHM}$ &   $V_{cent}$ & $V_{FWHM}$  \\ \tableline
2015-03-03        &  $-1.9\pm 0.4$   & $15.0\pm 0.3$    &  \nodata                  &  \nodata    \\ 
2017-03-22        &  $-0.3\pm 0.5$   & $15.8\pm 0.3$   & $-1.7\pm 0.5$  &  $13.4\pm 0.4$  \\
2020-02-15        &  $+1.3\pm 0.5$  & $15.0\pm 0.3$   & $-1.5\pm 0.5$  &  $13.8\pm 0.4$  \\
\tableline
\end{tabular}
\end{center}
\end{table}

\begin{figure}[th]
\begin{center}
\epsfig{file=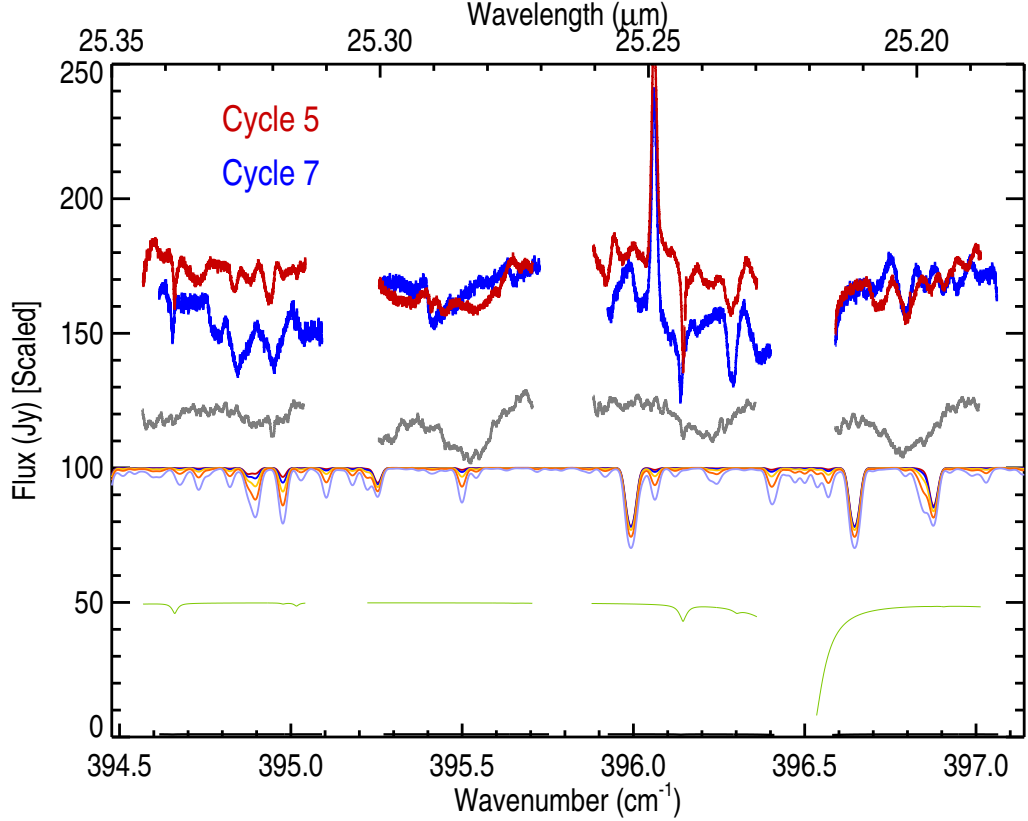, angle=90., width=15.0cm}
\caption{The spectral orders near the \siline line. The CY05 (red) and CY07 (blue) are uncorrected for order shape, and the CY05 Metis spectrum is shown in grey highlighting how the order shape can affect the stellar spectra. The most significant difference between CY05 and CY07 are the 
deepening of the strong water absorption lines in CY07 near $394.9\>{\rm cm}^{-1}$.
The bottom green line shows the sky spectrum.  The spherical MARCS model spectra, normalized
to 100 Jy, are described in Fig.~\ref{fig:orders_feii}. Again there is not a strong
correspondence between the photospheric models and the observed structure, in particular the 
feature on the high frequency side of the \siline emission.}
\label{fig:orders_s}
\end{center}
\end{figure}

\begin{figure}[th]
\begin{center}
\epsscale{0.85}
\epsfig{file=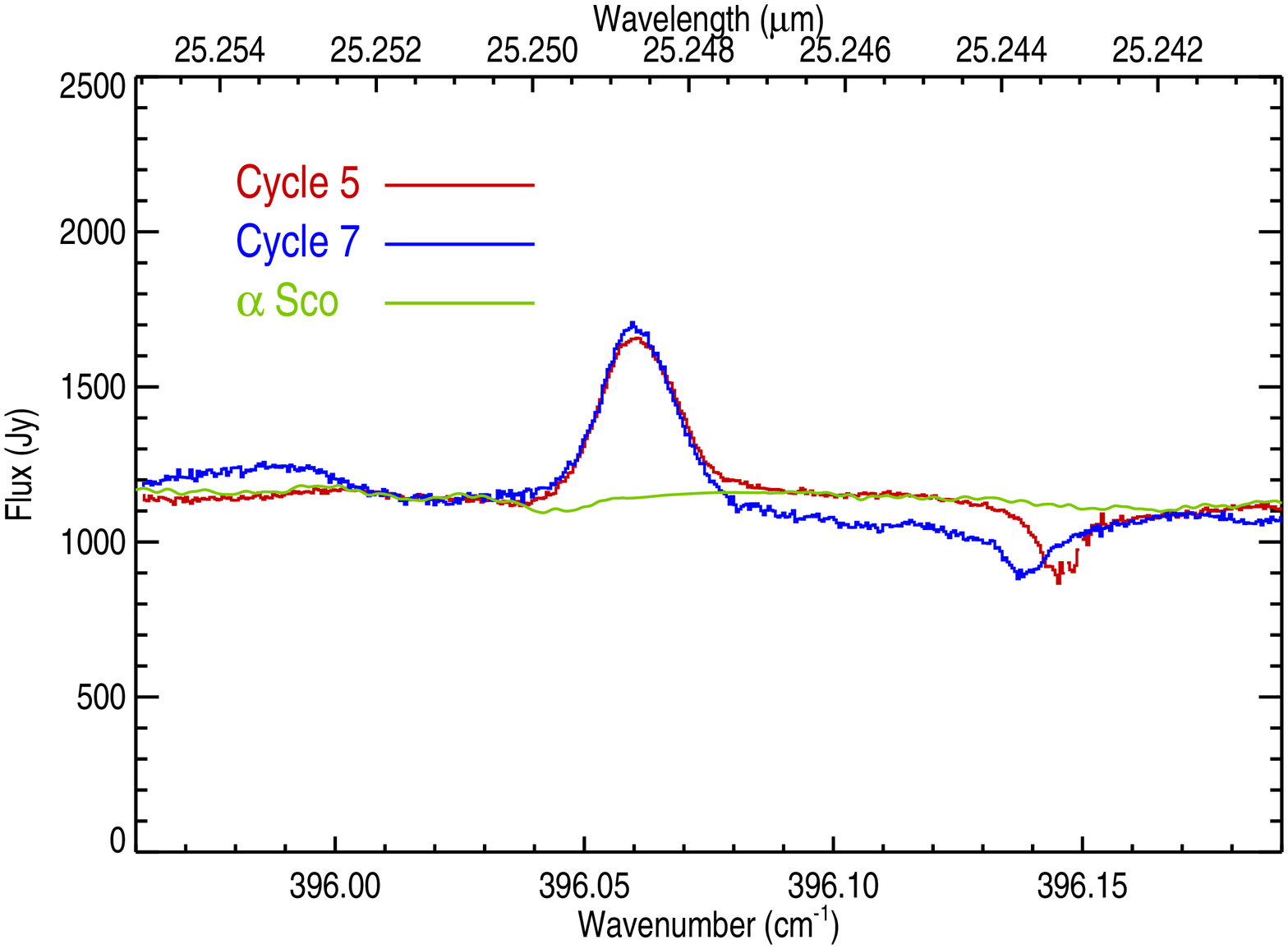, angle=90., width=13.5cm}
\caption{The Betelgeuse Cycle 5 and 7 \siline profiles obtained with the identical spectrograph setup.
The radial velocity Doppler shifts have been applied. Arbitrarily the continua have been normalized
at 396.30\,${\rm cm}^{-1}$. A previous observation of $\alpha$~Sco (M1.5~Iab) shows no \siline emission but its continuum near the emission line follows that of Betelgeuse in Cycle 5. $\alpha$~Sco's smoothed, and corrected for a telluric feature near 25.25\,$\mu$m, is shown in green. Like the \feiiline lines, the \siline lines show very similar line widths, centroid velocities, and similar flux to continuum ratios. Apart from changes in the continuum structure the circumstellar emission in the two lines at both epochs are very similar,
suggesting that the photospheric changes have not affected the region between $2-25R_\ast$ (See Discussion). Note the telluric features near $396.14\>{\rm cm}^{-1}$.}
\label{fig:si}
\end{center}
\end{figure}

The spectral orders near the \siline are shown in Fig.~\ref{fig:orders_s} and here there is a good
correspondence between the continuum features in CY05 and CY07. Again there is not a good agreement with the MARCS simulations. In CY07 the pair of strong water features near 
$394.9\>{\rm cm}^{-1}$ have deepened. Given that the continuum has a contribution from
silicate dust emission the features are much deeper than predicted by the coolest MARCS model,
with $T_{eff}=3200$\,K.

The two \siline emission profiles are shown in Fig.~\ref{fig:si} and it can be seen that these lines do not show a significant change in flux to continuum ratio either.  An examination of the \siline spectral region of $\alpha$~Sco (M1.5 Iab) obtained in Cycle 5 which shows no emission line, presumably a result of photoionization by the binary companion, shows a similar continuum to CY05. The 
\siline line measurements are given in Table~\ref{tab:results}. Again the dynamical properties are very similar between CY05 and CY07.  Note that the wavelength of the \feiiline line is quite accurately known, while the \siline line wavelength has an intrinsic uncertainty of 
$1\sigma(v)=2.4$\,\kms ~\citep{Blondel_2006JPhB...39.1409B}, so the value of the centroid velocities have a significant additional systematic uncertainty.

The \feiiline and \siline spectra reveal that the emission line properties have changed very little between the epochs, especially when the uncertainties in the profile measurements and the uncertain intrinsic continua are considered. It would seem contrived that the emission line and continuum fluxes would change in concert at different epochs, and with the similar line widths and velocity centroids it is reasonable to conclude that the CSE gas emitting these lines has not responded in a significant way to the changes that led to the 2020 February V-band minimum.  The CSE is not responding to either changes in the photosphere and/or to the potential presence of new dust in the sight-line to the star.

\subsection{\feiiline to \siline flux ratios }

We can also examine the \feiiline to \siline flux ratios for the 2017 and 2020 epochs by noting that the
continuum flux ratio for 25.25 to 25.99\,$\mu$m is expected to be close to unity, i.e., $\simeq 1.07$, based 
on the shape of the {\it ISO}-SWS spectra \citep{Justtanont_1999AA...345..605J, Sloan2003ApJS..147..379S}. This gives a \up[\ion{Fe}{2}\up] to \up[\ion{S}{1}\up] integrated line flux ratio of $\sim 15$, which
we can compare to theoretical predictions.

\citet{Carr_2000ApJ...530..307C} reported a near solar iron abundance for Betelgeuse [Fe/H]=$-0.02\pm 0.08$ [with $A({\rm S})_\odot=1.32\times 10^{-5}$ and $A({\rm Fe})_\odot=3.16\times 10^{-5}$ \citep{Asplund_2009ARA&A..47..481A}]. 
Both transitions are between the two lowest ground term fine-structure levels, and have
similar Einstein A-values (see Table~\ref{tab:atomic}). Neutral iron has a low first ionization potential 
(7.90\, eV) and is easily photoionized by the stellar ultraviolet continuum with $\lambda < 1570$\AA, while the second ionization potential is high (16.20\,eV) resulting in \ion{Fe}{2} being the dominant ionization state of the gas phase of the chromosphere and CSE outflow. Neutral sulfur has a high first ionization potential of 10.36\,eV with a ground state photoionization edge at 1195\AA{ }and requires far-ultraviolet flux to become photoionized. Sulfur is also expected to have a low association in any dust associated with the wind \citep{Snow_1987ApJ...321..921S}.

Both electron and neutral hydrogen collisions can drive a Boltzmann distribution for the low lying
fine-structure levels. While the electron density, $n_e$, is expected to be much lower than for 
hydrogen, $n_H$, it has larger collisional de-excitation rates.
Electron collision rates are available for \ion{S}{1} from \citet{Tayal_2004ApJS..153..581T} and \ion{Fe}{2} from \citet[][and Priv Comm]{Bautista_2015ApJ...808..174B},  S+H collision rates can be estimated using  the O+H rates of 
\citet{Lique_2018MNRAS.474.2313L}, and Fe+H rates are given by \citet{HM_1989ApJ...342..306H}. Using the thermodynamic models of  \citet{Rodgers_1991ApJ...382..606R},  \citet{Harper_2001ApJ...551.1073H},  or \citet{Harper_2017ApJ...836...22H} (where the gas temperatures are reduced close to the star)  both lines are formed predominantly above the critical densities (i.e., thermalized) when $T_{gas} > 500$\,K.  If \ion{Fe}{2} and \ion{S}{1} and are dominant ionization states, then the ratio of \feiiline and \siline fluxes,
would be $\le 4.5$ in the optically thin limit, and lower, $\sim 1.5$, when allowing for optical depth effects. The observations suggest then that the CSE sulfur is partially photoionized to \ion{S}{2} by the stellar chromosphere, where \ion{H}{1} Ly$\beta$ is the strongest source of line photons. The similarity of the flux ratios
between 2017 and 2020 also suggest that there has not been a strong change in the far-ultraviolet
radiation field, although if \ion{H}{1} Ly$\beta$ is a major photoionizing source then there will be a time-lag for the photons in this very opaque line to escape to the CSE.

\section{Discussion\label{section:discussion}}

The EXES observations show no significant changes in the velocity centroids and line widths between 2015 and 2017, and 2020 February  when the V-band brightness reached its minimum state. 
The line to continuum ratios also show no evidence that the wind heating has changed. The event leading to the great dimming in V-band has not significantly  affected the inner circumstellar outflow.
Magnetic fields have long been considered a candidate for driving the outflows from RSG
\citep{HA_1984ApJ...284..238H}, and Betelgeuse's measured magnetic field 
\citep{Auriere2010AA...516L...2A}  appears typical of other RSGs 
\citep{Tessore2017AA...603A}. The time-scale for photospheric magnetic field variations 
\citep{Mathias_2018A&A...615A.116M} is consistent with the 5.8\,yr time-scale observed in radial velocity and V magnitudes which has been attributed to giant convective cells.
If this is the case, the large-scale morphology of the open magnetic field lines might be modified by the emergence of a new large cool convective cell, one might expect that the Alf\'ven wave properties
and wind acceleration might respond to such a change.

The thermodynamic models of \citet{Harper_2017ApJ...836...22H} suggest that the
\feiiline and \siline lines are predominantly formed within $3-20R_\ast$. Assuming a distance of 222\,pc \citep{Harper_2017AJ....154...11H} this corresponds to a light crossing time of a few hours.  The radius of the peak mean chromospheric temperature derived from radio continuum observations is $\simeq 1.75R_\ast$ \citep{Eamon_2015A&A...580A.101O}, and if the atmospheric disruption has disturbed the magnetic field in this region then the time for outward propagating Alfv\'en waves to cross this region can be estimated. Using the thermodynamic model of \citet{Harper_2001ApJ...551.1073H}, and assuming a radial magnetic field with $B(1.75R_\ast) \sim 5$\,G, a typical value expected on energetic grounds
\citep[e.g.,][]{HA_1984ApJ...284..238H}, the Alfv\'en wave-crossing time to 
$10R_\ast $ is of the order of a decade.
The absence of changes in the CSE emission in this region is consistent with photospheric changes in
large-scale magnetic fields that have not yet reached the inner-CSE.

The pattern of V magnitude and \up[O~I\up] fluxes noted by \citet{Haas1995ASPC...73..397H},
if confirmed, might then result from a change in wind heating  caused by gas-dust grain collisions in a region where $T_{gas} < 500$\,K.  
\citet{Haas1995ASPC...73..397H} suggest that the \oiline fluxes might be related to V-band brightness changes through a change in the dust grain drift velocity \citep{GS_1976ApJ...205..144G}.
The \oiline has flux contributions that overlap the \feiiline and \siline lines but also extends farther out to $\sim 3$\arcsec ($\sim 130R_\ast$).

The theoretical study of the CSE's thermal steady-state structure by \citet{Rodgers_1991ApJ...382..606R} showed that the onset of CSE dust heating, near $30R_\ast$ 
is matched by a peak in \oiline cooling.  Dust can respond quickly to changes in the photospheric illumination, and the dust-gas heating rate per unit volume $H$ is proportional to the cube of the dust-drift velocity, or to changes in effective luminosity in the {\it steady state}  as $H \propto L^{3/2}$. 
However, there also will be a temporal lag between changes in the luminosity and heating of the gas.  
While the V-mag has increased by $\Delta V\simeq 1$-mag since 2009 September, the 
narrow-band Wing $1.0240\>\mu$m C-band photometry reveals a smaller increase of $\Delta C\simeq 0.3$-mag, so it is not clear what the magnitude of the changes in dust properties are likely to be. However, assuming a radiation pressure efficiency of $Q_{Pr}=0.5$, a grain radius of 0.1$\mu$m, and gas densities from CSE models, the periodic dust-drift velocity lags the luminosity variations by $\sim 20$\,days. The collisional excitation and  radiative decay time-scales for the \oiline are about a week so it is possible that, in regions where dust heating dominates the CSE energy balance, changes in stellar luminosity can modify the gas temperature on times-scales much shorter than Betelgeuse's 420\,day period.

Reducing the dust heating in the CSE gas will only reduce the emission lines fluxes if its magnitude is comparable to other
heating and cooling processes.  For the \feiiline and \siline lines, the dust plumes seen at similar formation radii in images, may be such an example, or instead the dust may represent a small volume filling factor.

\section{Conclusions\label{section:conclusions}}

These EXES results, which do not reveal significant changes in the \feiiline and \siline emission lines between 2017 March and 2020 February, when combined with contemporaneous and future multi-wavelength observations will help to clarify the nature of the 2019-2020 V-band dimming of Betelgeuse and its subsequent effect on the extended atmosphere. The EXES results suggest that dust in the inner-CSE
is not significantly heating the gas. 
At the time of submission of this Letter \sofias will be observing Betelgeuse with the  Far Infrared Field-Imaging Line Spectrometer (FIFI-LS)  \citep{FIFI_Colditz_2018JAI.....740004C, FIFI_2018JAI.....740004C}, upGREAT 
\citep{upGREAT_2018JAI}.

\bibliography{harper_DDT_2020Feb16}

\acknowledgments

GMH's research was supported in part by SOFIA grants SOF~$05\_0073$ and SOF~$07\_0073$.
This research has made extensive use of NASA's Astrophysics Data System. 
EXES is currently supported through NASA-UC~Davis collaborative agreement No. 
80NSSC19K1701. We thank the referee for helpful suggestions that improved this Letter.

\vspace{5mm}
\facilities{ISO,KAO,SOFIA(EXES,GREAT,FIFI-LS)}

\end{document}